\documentclass[12pt]{article}

\usepackage{framed}
\usepackage{cancel}
\usepackage{braket}
\usepackage{graphicx}
\usepackage{mathrsfs}
\usepackage{amsmath,amssymb}
\usepackage{listings}
\usepackage[most]{tcolorbox}
\usepackage{inconsolata}
\usepackage{times}
\usepackage{hyperref}
\usepackage[english]{babel}
\usepackage[utf8]{inputenc}
\usepackage{cancel}
\usepackage[normalem]{ulem}
\usepackage{stmaryrd}

\usepackage{soul} 
\usepackage{xcolor} 

\newcommand{\floor}[1]{\left \lfloor #1 \right \rfloor}
\newcommand{\jump}[1]{\left [ #1 \right ]}
\newcommand{\shell}{\partial \mathcal{M}}
\newcommand{\smet}{h}

\newtcblisting[auto counter]{codeprint}[2][]{sharp corners,
    fonttitle=\bfseries, colframe=gray, listing only,
    listing options={basicstyle=\ttfamily,language=C++},
    title=C\'odigo \thetcbcounter: #2, #1}

\numberwithin{equation}{section}

\newcounter{lastnote}

\definecolor{blue-violet}{rgb}{0.54, 0.17, 0.89}
\definecolor{PineGreen}{cmyk}{0.92, 0, 0.59, 0.25}
\definecolor{Gray}{cmyk}{0, 0, 0, 0.50}

\title{Thin shell dynamics in Lovelock gravity}
\author{Pablo Guilleminot$^a$, Nelson Merino$^b$\\
  and Rodrigo Olea$^a$ \smallskip \\
$^a${\small \emph{Departamento de Ciencias F\'isicas, Universidad
Andres Bello,}}\\
{\small \emph{Sazi\'e 2212, Piso 7, Santiago, Chile. \smallskip }}\\
$^b${\small \emph{Instituto de Ciencias Exactas y Naturales (ICEN), Facultad de Ciencias,}}\\
{\small \emph{ \smallskip  Universidad Arturo Prat, Iquique, Chile}}}

\date{\today}
\begin{document}

\baselineskip20pt

\maketitle

\begin{abstract} 
We study matching conditions for a  spherically symmetric thin shell in Lovelock gravity which can be read off from the variation of the corresponding first-order action. In point of fact,  the addition of Myers´ boundary terms to the gravitational action eliminates the dependence on the \emph{acceleration} in this functional and such that the canonical momentum appears in the surface term in the variation of the total action. This procedure leads to junction conditions  given by the discontinuity of the canonical momentum defined for an evolution normal to the boundary.\\                                        
In particular, we correct existing results in the literature for the thin shell collapse in generic Lovelock theories, which were mistakenly drawn from an inaccurate analysis of the total derivative terms in the system.

\end{abstract}

\section{Introduction}
Junction conditions describe the behavior of physical  fields across surfaces where matter density is discontinuous. A simple example is found in electromagnetism: the integration of the Maxwell equations over a pillbox, which encloses a charge/current density, leads to a jump in the electric/magnetic field.

Junction conditions also play an essential role in gravitational collapse dynamics. In the case of Snyder-Oppenheimer collapse \cite{Oppen}, one deals with a matter-vacuum interface where the density is given by a step function defined at the star boundary.  By contrast, when treating thin shells, the source is localized by means of a stress tensor with a Dirac delta distribution. This allows to develop the analysis of spherically symmetric thin shells, studied in the theory of General Relativity (GR) by W. Israel \cite{Isra}. The projection of the field equations to the shell frame, upon imposing continuity of the spacetime metric, enables to express the jump as a precise combination of the extrinsic curvature and its trace, sourced by the matter on the shell.

An important application of junction conditions is \textit{braneworld} scenario \cite{Shiro}, where the dynamics of the universe itself is described as a four-dimensional brane embedded in a five-dimensional bulk spacetime.
The source is an energy-momentum tensor proportional to a delta function. In order to balance the presence of this function, a discontinuity appears on the other side of the equation. Thus, the dynamics of the brane is given by the Israel matching conditions.

Junction conditions have been also obtained for gravity theories which represent modifications to GR. In higher-derivative gravity, junction conditions were first studied in Refs.\cite{Parry:2005eb,Nojiri:2001ae}.
Later, it was shown that different types of junction conditions can be obtained for a general quadratic theory of gravity, depending on how singular the metric is allowed to be \cite{Deruelle:2007pt}. The treatment is carried out in a similar way as in GR, by identifying the term with the highest normal derivative. 

On the other hand, matching conditions can also be derived from the variational principle worked out in terms of the discontinuity of the canonical momentum of the theory. In order to make this manifest, the example of electromagnetism is revisited. In Minkowski space the dynamics is governed by the Maxwell action
\begin{equation} \label{EMaction}
  I=\int\limits_{\mathcal{M}}d^4x\,\left (\frac{1}{4}F^{\mu\nu}F_{\mu\nu}+A_\mu J^\mu \right)\,,
\end{equation}
where $F_{\mu\nu}=\partial_\mu A_\nu-\partial_\nu A_\mu\,$ is the Faraday tensor and $A_\mu=A_\mu(x)$ is the electromagnetic four-potential minimally coupled to a four-current density $J^\mu$. 
Notice also that due to the dependence  on quadratic terms in the first derivative of the fields, the variational principle for the action (\ref{EMaction}) is satisfied by Dirichlet boundary conditions. Indeed, the variation of the above equation gives
\begin{equation}
\delta I=\int\limits_{\mathcal{M}}\left(-\partial_{\mu}F^{\mu\nu}+J^{\nu}\right)  \delta
A_{\nu}+\int\limits_{\partial\mathcal{M}} n_{\mu} F^{\mu\nu}\delta A_{\nu}    
\end{equation}
Now, consider a localized source in a sheet $\partial\mathcal{M}$ which, in Cartesian coordinates $x^{\mu}=(t,x,y,z)$, is placed at $z=0$. The manifold is divided in two regions $\mathcal{M}_{+}$ and $\mathcal{M}_{-}$, which share the boundary $\partial\mathcal{M}$ with the normal vector $n_\mu=\delta_\mu^z$ and current density $J^\mu=j^\mu \delta(z)$. 
In this treatment, it is implicit the assumption that the transversal components of the gauge connection $A_{i}$ (where the Latin indices label the coordinates $x^{i}=(x,y,t)$) are continuous across the shell. As the current density is given by a delta function, the integration in the normal direction can be performed, i.e., \emph{moving} the source to the interface between the two regions. The variation of the action leads, on-shell, to a surface term of the form 

\begin{equation}\label{vos}
  \delta I=\int\limits_{\partial\mathcal{M}}d^3x\,\left (-\jump{F^{zi}}+j^i \right)\delta A_i\,,
\end{equation}
where $\jump{F}=F^{+}-F^{-}$ refers to the difference of a quantity $F$ between $\mathcal{M}_{+}$ and $\mathcal{M}_{-}$. Since $\delta A_i$ is arbitrary at the interface, the variational principle $\delta I=0$ implies the matching condition 
\begin{equation}
\jump{F^{zi}}=j^i\,.
\end{equation}
It is straightforward to see that the last equation reproduce the known junction conditions
\begin{equation*}
\vec{n}\cdot\jump{\vec{E}}=\sigma \quad , \quad  \vec{n}\times\jump{ \vec{B}}=\vec{j}\,,
\end{equation*}
for a generic normal vector to the boundary.
Furthermore, for an evolution of the system along the normal coordinate $z$ the associated canonical momentum is 
\begin{eqnarray}
\pi^i=\frac{\partial \mathcal{L}}{\partial \big(\partial_z A_i\big)}=F^{z i}\,.
\end{eqnarray}
From this simple example, one can understand the analysis of junction conditions as coming from the discontinuity of the canonical momentum (associated to the normal evolution).\footnote{In a more contemporary context, in axion electrodynamics, where the Maxwell Lagrangian is augmented by a pseudo-scalar field coupled to the Pontryagin term for $U(1)$, the corresponding junction conditions can also be derived in this way \cite{Omar}. }


It is an appealing idea to extend this procedure to obtain junction conditions for GR. However, in contrast with the Maxwell Lagrangian which is quadratic in first derivative terms, the Einstein-Hilbert (EH) Lagrangian depends on second order derivatives. 
As a consequence, the variation of the EH action involves variations of the metric and its first derivative. 
In order to have a well-posed  variational principle for Dirichlet boundary conditions, the term which is linear in the normal acceleration must be
eliminated by the addition of the Gibbons-Hawking-York (GHY) boundary term. As a result, a first-order Lagrangian density is obtained and can be used to derive matching conditions for thin shells in gravity. This statement will be revised in the following sections.

The derivation of the junction conditions in terms of the canonical momentum is not clear in higher-derivative gravity.
On one hand, it is expected that higher order momenta would exist in that class of theories. On the other hand, different matching conditions can arise according to how singular the metric is considered. Thus, from the point of view of the variational problem, there is no unambiguous interpretation of the quantity that should jump across the shell.

For these reasons, in this work we focus on Lovelock gravity: the most general covariant theory in a $D$-dimensional spacetime giving second order divergenceless field equations \cite{LL}.
In presence of a boundary, akin to General Relativity, the variation the Lovelock actions produce boundary terms that depend on variations of the metric and its first derivatives. Consequently, a well-posed variational principle -- for a Dirichlet condition on the boundary metric-- requires the Lovelock action to be supplemented by Myers terms \cite{Myers}. 

The above idea was used in Ref.\cite{Davis} (see also Refs. \cite{Charmousis:2002rc,Gravanis:2002wy,GaGiGraWi,Lin:2010wy}) to derive junction conditions for an Einstein-Gauss-Bonnet (EGB) braneworld. Knowing that Israel matching  conditions can be derived by varying the EH action with the GHY term, they obtained generalized junction conditions for EGB theory from the variational principle with the corresponding Myers/Müller-Hoissen term \cite{MuellerHoissen:1989yv}.
Later, in Ref.\cite{OlRo}, it was shown that the variation of a Lovelock density plus its corresponding Myers term reproduces the Hamiltonian variation made in Ref.\cite{TeiZan}.
In particular, the canonical momentum can be easily read off from the surface term of the variation of the Dirichlet action (see also Chapter 15 of Ref.\cite{Padmanabhan:2010zzb}). It is then clear that the canonical momentum plays the role of a generalized Brown-York stress-tensor \cite{BroYo} in Lovelock gravity. 
Therefore, it makes sense to study generalized junction conditions for a generic Lovelock action as the discontinuity in the canonical momenta where the brane itself is a boundary . This problem will be addressed in detail in this work.

As stressed in Ref.\cite{GJMO}, the Dirichlet action is a first-order  functional whose variation coincides with the  one of the Hamiltonian action \cite{TeiZan}. As a matter of fact, both actions produce a surface term of the form $\pi^{ij} \delta h_{ij}$.
As the corresponding first-order Lagrangian ${\cal{L}}_{1}$ is the one which enters in the definition of Hamiltonian, it is the proper function to which the canonical momentum is associated to.

As we shall see below, the variational principle and the use of adapted coordinates on the shell, allows to work out the shell dynamics in Lovelock gravity, as coming from the jump in the canonical momentum across it.
In particular, we correct existing results in the literature, which are a consequence of a wrong analysis of junction conditions in Lovelock theory \cite{CSD}.

This work is organized as follows: In section \ref{sec1}, we briefly review the Dirichlet problem for Einstein-Hilbert, Einstein-Gauss-Bonnet and Lovelock theories in the Gauss-normal coordinate frame. In the process, the connection between first-order Lagrangians and the Dirichlet problem is revisited. With  this in mind, we show in section \ref{sec2} how the variational principle gives rise to junction conditions for  a thin shell. Then, in section \ref{sec3}, we carry out the procedure described above to obtain explicit expressions for the junction conditions for Lovelock gravity. In the Conclusions, we summarize our results and explore future directions.

\section{The Dirichlet action for Lovelock gravity} \label{sec1}

In what follows, we make extensive use of Gauss-normal coordinates
\begin{equation} \label{GN}
ds^2=g_{AB} dx^A dx^B=N^2(w)dw^2+h_{ab}(w,x)dx^adx^b\,,
\end{equation}
where the spacetime is foliated as an infinite series of surfaces of constant $w$ with normal $n_A=N\delta^w_A$, each one with an induced metric $h_{ab}$. In this frame, we consider the splitting of the spacetime indices as $A=(w,a)$, such that the different components of the Riemann curvature tensor are given by the Gauss-Codazzi-Mainardi (GCM) relations in Appendix \ref{GC}. Furthermore, the normal derivative $\partial_n$, defined for the vector which produces the spacetime foliation (\ref{GN}), is expressed as
\begin{equation} \label{partn}
\partial_n=n^A\partial_A=\frac{1}{N}\partial_w\,.
\end{equation}

\subsection{Einstein gravity}

In General Relativity, the dynamics is described by the Einstein-Hilbert action in $D=d+1$ dimensions 
\begin{equation}\label{EH}
I_{\rm EH}=\frac{1}{16\pi G}\int\limits_{\mathcal M}d^Dx\sqrt{-g}\big(R-2\Lambda\big)\ ,
\end{equation}
where $R$ is the Ricci scalar, $g=\det g_{AB}$ is the metric determinant and $\Lambda$ is the cosmological constant.
A well-posed action principle is achieved if the Einstein-Hilbert action is supplemented by the Gibbons-Hawking-York (GHY) term \cite{York:1972sj,Gibbons:1976ue}
\begin{equation}\label{ID}
I[g]=\frac{1}{16 \pi G}\int\limits_{\mathcal{M}}d^Dx\sqrt{-g}\,\big(R-2\Lambda\big) -\frac{1}{8 \pi G}\int\limits_{\partial\mathcal{M}}d^dx\sqrt{-h}K\,,
\end{equation}
where $K$ is the trace of the extrinsic curvature $K_{ab}=-\frac{1}{2N} \partial_wh_{ab}$, where a prime stands for $\partial_w$.
Arbitrary variations of this action give
\begin{equation} \label{IDvar}
\delta I=-\frac{1}{16\pi G}\int\limits_{\mathcal{M}}d^D x\,\mathcal{E}^{AB}\delta g_{AB}+\frac{1}{16\pi G}\int\limits_{\partial\mathcal{M}}d^d x%
\,\sqrt{-h}\pi^{ab}\delta h_{ab}\,,
\end{equation}
where the bulk expression is $\mathcal{E}_{AB}=G_{AB}+\Lambda g_{AB}$ --with $G_{AB}=R_{AB}-\frac{1}{2}R\,g_{AB}$ the Einstein tensor-- and the boundary stress tensor, which can be read off from the surface term in the above equation, is
\begin{equation} \label{pi1}
\pi^{ab}=K^{ab}-Kh^{ab}\,.
\end{equation}
The Dirichlet action (\ref{ID}) yields a well-defined variational problem, 
as it remains stationary when the metric is kept fixed at the boundary. As a matter of fact, it is manifest that its variation (\ref{IDvar}) vanishes when imposing the boundary condition 
$\delta h_{ab}=0$ at ${\partial\mathcal M}$.

One can use Gaussian coordinates (\ref{GN}) and the GCM relations (\ref{riemann}) in order to lift the GHY boundary term to the bulk, i.e., to express it as a part of the bulk Lagrangian. In doing so, one can show that the Lagrangian density associated to the action (\ref{ID}) is equal to  
\begin{align}
\mathcal L_{\rm 1}&=N\sqrt{-h}\bigg(\bar R(h)+K^2-K^{ab}K_{ab}-2\Lambda\bigg)\,.\label{L1_GR}
\end{align}
where the bar stands for a quantity computed with the boundary metric. This is the time-honored, first-order Arnowitt, Deser and Misner (ADM) Lagrangian of Einstein gravity \cite{ADM}, essential ingredient to construct a Hamiltonian in Einstein gravity. As a direct consequence of this fact, the tensor (\ref{pi1}) can be identified with the canonical momentum, conjugate to the dynamic variable $h_{ab}$, i.e.,
\begin{align}
\pi^{ab}&=\frac{1}{\sqrt{-h}} \frac{\partial \mathcal L_{1}}{\partial(\partial_w h_{ab})}\,.\label{eqDiffLagrangianGR}
\end{align}
One may recover the full generality of the procedure, beyond the particular gauge choice (\ref{GN}). Indeed, in the general ADM decomposition, the extrinsic curvature $K_{ab}$ is extended to be
\begin{equation}
K_{ab}=-\frac{1}{2N(w)}\left(h'_{ab}-\nabla_aN_b-\nabla_bN_a\right).
\end{equation}
Upon integrating by parts, a term proportional to $N^a\bar{\nabla}_b\pi^b_a$ is added to the Lagrangian $\mathcal L_1$ considered above. It can be explicitly checked that variations with respect to $N$, $N^a$, and $h_{ab}$ of $\mathcal{L}_{1}$ yield, respectively, the constraints $\mathcal{E}^w_w=0$ and $\mathcal{E}^w_a=0$, and the dynamical component $\mathcal{E}^a_b=0$ of the equations of motion
\begin{eqnarray}
\mathcal{E}^w_w & = & -\frac{1}{2}\left(\bar{R}-K^2+K^a_bK^b_a\right)+\Lambda \label{eom1}\\
\mathcal{E}^w_a & = &  \frac{1}{N}\left (\bar{\nabla}_a K - \bar{\nabla}_c K^c_a\right ) \label{eom2}\\
\mathcal{E}^{a}_b & = & \bar{\mathcal{G}}^{a}_b+\Lambda\delta^a_b-K\big(K^a_b-K\delta^a_b\big)+\frac{1}{2}\big(K^c_dK^d_c-K^2\big)\delta^a_b+\partial_n(K^a_b-K\delta^a_b) \label{eom3}\,.
\end{eqnarray}
where $\partial_n$ is given by Eq.(\ref{partn}).

For the above action, endowed with the Gibbons-Hawking boundary term, Einstein equation (\ref{eom2}) implies the conservation of the canonical momentum (\ref{canmom}). For a given set of Killing vectors $\{\xi^{i}\}$ the corresponding conserved quantities were derived in Ref.\cite{BroYo}, which are now broadly known as Brown-York charges
\begin{equation} \label{BJCC}
Q[\xi]=-\frac{1}{16\pi G}\int\limits_{\Sigma} d^{D-2}x\sqrt{\sigma} \,u_{a}\left(\pi^{a}_{b}-\pi_{[0]b}^{a}\right)\xi^{b}\,.\notag \\
\end{equation}
Here, $u_a$ is the normal to the co-dimension 2 surface $\Sigma$ and $\sigma$, the determinant of its metric, typically expressed in Schwarzschild-like coordinates. As this is a background-dependent notion of energy and other conserved charges, the subscript $[0]$ in the momentum denotes evaluation on the corresponding vacuum geometry, e.g., Minkowski or global de Sitter (dS) or anti-de Sitter (AdS) spacetimes. 

\subsection{Einstein-Gauss-Bonnet gravity}

An arbitrary modification of General Relativity to include quadratic-curvature couplings in the action leads, in general, to fourth-order field equations. On the contrary, only for a precise combination of such terms --known as Gauss-Bonnet-- the EOM are still of second order in derivatives of the metric. The action for Einstein-Gauss-Bonnet gravity is 

\begin{equation}\label{EGB}
I_{\rm EGB}=\frac{1}{16\pi G}\int\limits_{\mathcal M}d^Dx\sqrt{-g}\left (R-2\Lambda+\alpha\Big(R^{AB}_{CD}R^{CD}_{AB}-4R^{AB}R_{AB}+R^2\Big)\right )\ ,
\end{equation}
where $\alpha$ is the Gauss-Bonnet coupling. In order to turn this action compatible with Dirichlet boundary conditions, it is necessary to supplement it with a generalization of the Gibbons-Hawking term, i.e.,

\begin{eqnarray} \label{EGBD}
&& I =\frac{1}{16\pi G}\int\limits_{\mathcal M}d^Dx\sqrt{-g}\left (R-2\Lambda+\frac{\alpha}{4}\delta^{[A_1\cdots A_{4}]}_{[B_1\cdots B_{4}]}R^{B_1B_2}_{A_1A_2}%
R^{B_{3}B_{4}}_{A_{3}A_{4}}\right )\hspace{50pt}\notag \\
&&\hspace{50pt}-\frac{1}{8 \pi G}\int\limits_{\partial\mathcal{M}}d^dx\sqrt{-h}\left (K+\alpha\delta_{[b_1\cdots b_{3}]}^{[a_1\cdots a_{3}]}K^{b_1}_{a_1}\left( \frac{1}{2}\bar{R}^{b_2b_3}_{a_2a_3}-\frac{1}{3}K^{b_2}_{a_2}K^{b_3}_{a_3} \right)\right )\,.
\end{eqnarray}
For conventions on generalized Kronecker deltas of higher rank, see Appendix \ref{DK}. The variation of the above action takes the form
\begin{equation} \label{IDvarGB}
\delta I=-\frac{1}{16\pi G}\int\limits_{\mathcal{M}}d^D x\,\mathcal{E}^{AB}\delta g_{AB}+\frac{1}{16\pi G}\int\limits_{\partial\mathcal{M}}d^d x%
\,\sqrt{-h}\,\pi^{ab}\delta h_{ab}\,,
\end{equation}
where, in this case, the equations of motion are 
\begin{equation}
\mathcal{E}^A_B=G_{B}^A+\Lambda\delta^A_B -\frac{\alpha}{8}\delta^{[AA_1\cdots A_{4}]}_{
[BB_1\cdots B_{4}]}%
R^{B_1B_2}_{A_1A_2}R^{B_{3}B_{4}}_{A_{3}A_{4}}=0\,.
\end{equation}
and the tensor $\pi^{ab}$ in the boundary term is 
\begin{eqnarray} \label{Pi}
\pi^a_b=K^a_b-K\delta^a_b-2\alpha%
\delta^{[aa_1\cdots a_{3}]}_{[bb_1\cdots b_{3}]}K^{b_1}_{a_1}\left( \frac{1}{2}\bar{R}^{b_2b_3}_{a_2a_3}-\frac{1}{3}K^{b_2}_{a_2}K^{b_3}_{a_3} \right)\,.
\end{eqnarray}
It is worthwhile to notice that the index structure of the part of $\pi^a_b$  associated to the Gauss-Bonnet term makes apparent the fact the canonical momentum vanishes identically in $D=4$. Therefore, it is clear that the Dirichlet problem for the metric $h_{ab}$ cannot be defined in the case of the addition of the Gauss-Bonnet term to the four-dimensional gravitational action\footnote{Only the proper use of asymptotic conditions in anti-de Sitter gravity leads to a consistent Dirichlet problem in 4D, but for the holographic metric at the conformal boundary instead \cite{AMOP}.}.

As in the GR case, the use of Gaussian coordinates and the GCM relations allows to \emph{lift} the boundary term to the whole spacetime, such that it now appears as a bulk term. In doing so, the second normal derivatives of the metric are eliminated from the bulk Lagrangian density \cite{GJMO}, and the action (\ref{EGBD}) is then written as 
\begin{eqnarray*}
\mathcal{L}_{\rm 1}&=&N\sqrt{-h}\Bigg[%
\bar R+K^2-K^{ab}K_{ab}-2\Lambda+\frac{\alpha}{4}\delta_{\lbrack b_{1}\cdots b_{4}%
]}^{[a_{1}\cdots a_{4}]}\bar R_{a_{1}a_{2}}^{b_{1}b_{2}}\bar R_{a_{3}a_{4}}^{b_{3}b_{4}}+\hspace{100pt}\notag \\
&&\hspace{150pt}+\alpha\delta_{\lbrack b_{1}\cdots b_{4}]}^{[a_{1}\cdots a_{4}]}K_{a_{1}}^{b_{1}}K_{a_{2}}^{b_{2}}\left(\bar R_{a_{3}a_{4}}^{b_{3}b_{4}}-\frac{1}{3}K_{a_{3}}^{b_{3}}K_{a_{4}}^{b_{4}}\right)\Bigg]\,. \end{eqnarray*}
This is a first-order Lagrangian density, which depends on the first normal derivative of the induced metric and allows to identify the tensor (\ref{Pi}) as the canonical momentum conjugate to $h_{ab}$.

The components $\mathcal{E}^w_{w}$ and $\mathcal{E}^w_a$ of the equations of motion for Einstein-Gauss-Bonnet gravity take the form
\begin{equation}
\mathcal{E}^w_w = G^w_w+\Lambda-\frac{\alpha}{8}\delta_{[b_{1}\cdots b_{4}%
]}^{[a_{1}\cdots a_{4}]}R_{a_{1}b_{2}}^{b_{1}b_{2}}R_{a_{3}a_{4}}^{b_{3}b_{4}} \quad ,\quad \mathcal{E}^w_a=-\frac{1}{N}\bar{\nabla}_b \pi^b_a
\end{equation}

In addition, the component $\mathcal{E}^a_{b}$ is given by
\begin{eqnarray*} 
\mathcal{E}^a_{b}&=&\partial_n\pi^a_b-\delta^{[aa_1\cdots a_{4}]}_{[bb_1\cdots b_{4}]}%
K^{b_1}_{a_1}K^{b_2}_{a_2}\Bigg( \frac{1}{(D-3)(D-4)}\delta^{b_3}_{a_3}\delta^{b_4}_{a_4}+\frac{\alpha}{2}\bar R^{b_3b_4}_{a_3a_4}- \frac{\alpha}{6}K^{b_3}_{a_3}K^{b_4}_{a_4} \Bigg)+ \notag \\
&&\hspace{50pt}+\left(K^a_c-K\delta^a_c-2\alpha%
\delta^{[aa_1\cdots a_{3}]}_{[cb_1\cdots b_{3}]}K^{b_1}_{a_1}\left( \frac{1}{2}\bar{R}^{b_2b_3}_{a_2a_3}-\frac{1}{3}K^{b_2}_{a_2}K^{b_3}_{a_3} \right)\right)\left(K^c_b-K\delta^c_b\right) \notag \\
&&\hspace{190pt}+\bar{\mathcal{E}}^a_{b}-\frac{\alpha}{2}\delta_{[ bb_{1}b_{2} b_{3}]}^{[aa_1a_{2} a_{3}]}\bar{\nabla}_{a_1}\bigg(K_{a_{2}}^{b_{2}}%
\bar{\nabla}^{b_{1}}K_{a_{3}}^{b_{3}}\bigg)+\Lambda\delta^a_b\,.
\end{eqnarray*}
As in the case of General Relativity, the field equation $\mathcal{E}^w_ {a}=0$ implies a conservation law written in terms of the tensor $\pi^{ab}$. Notwithstanding the foregoing, this tensorial quantity is identified with the canonical momentum only if the corresponding Lagrangian for Einstein-Gauss-Bonnet gravity is of first-order in normal derivatives. In adapted, Gaussian coordinates, the derivation of the canonical momentum in GR makes it equivalent to the Brown-York stress tensor \cite{BroYo}. The extension of the notion of Brown-York energy-momentum tensor to EGB gravity is then naturally realized by Eq.(\ref{Pi}). However, the fact $\pi^{ab}$ correctly accounts for the the energy of, e.g., Boulware-Deser black holes \cite{BoulDer}, would depend on the value of the cosmological constant. For asymptotically flat solutions, the mass obtained is the correct one. In turn, for asymptotically AdS black holes, only a fraction of the mass is obtained, with a factor which depends both on the dimension and the GB coupling. The addition of local counterterms at the boundary happens to correct the factor and to remove infrared divergences at radial infinity \cite{BriRa,LiuSab}.

\subsection{Lovelock gravity}

Lovelock gravity is the natural generalization of Einstein theory for $D>4$.  Evidence hinting at the appearance of such terms have been found in the low energy effective action of heterotic string theory \cite{bh01,bh02,bh03} and six-dimensional Calabi-Yau compactifications of M-theory \cite{bh04}. This fact brought a thirst for solutions of this model. Besides Boulware-Deser black hole solution, Wiltshire solved the EGB system with the Maxwell term included \cite{bh05}. In turn, in Ref.\cite{bh08} the Born-Infeld model was studied. By dropping the spherically symmetric condition, topological black holes were found \cite{bh07}, featuring flat and hyperbolic transversal sections. 
Beyond the Gauss-Bonnet term, spherically symmetric black holes were discovered in maximally degenerate Lovelock gravity \cite{bh10}. In Ref.\cite{BHS}, such solutions were extended to Lovelock Unique Vacuum (LUV) theories with intermediate multiplicity. Also, black holes were explored 
for Lagrangians containing a single Lovelock term plus cosmological constant \cite{bh09}.

The goal of this section is to review the construction of the first-order Lagrangian for Lovelock gravity. 

Consider the Dirichlet action of a generic
Lovelock theory
\begin{equation} \label{IDL}
I=\frac{1}{16\pi G}\sum_{p=0}^{\floor{\frac{D-1}{2}}}\alpha_p\Bigg(\int\limits_{\mathcal{M}}d^Dx\mathcal L^{(p)}-%
\int\limits_{\partial\mathcal{M}}d^{d}x\beta^{(p)}\Bigg)\,,
\end{equation}

where $\floor{\,\cdot\,}$ is the floor function and ${\alpha_{p}}$ is a set of arbitrary coupling constants. The term of degree $p$ in the curvature is
\begin{equation} \label{Lp}
\mathcal L^{(p)}=\frac{1}{2^p}\sqrt{-g}\delta^{[A_1\cdots A_{2p}]}_{[B_1\cdots B_{2p}]}R^{B_1B_2}_{A_1A_2}\cdots%
R^{B_{2p-1}B_{2p}}_{A_{2p-1}A_{2p}}\,,
\end{equation}
that has the property to be topological in $D=2p$ dimensions. Its corresponding Myers term 
\begin{eqnarray} \label{B2n}
\beta^{(p)}&=&2p\sqrt{-h}\int_0^1ds\,\delta^{[a_1\cdots a_{2p-1}]}_{[b_1\cdots b_{2p-1}]}K^{b_1}_{a_1}\left( \frac{1}{2}\bar{R}^{b_2b_3}_{a_2a_3}-s^2K^{b_2}_{a_2}K^{b_3}_{a_3} \right)\times \cdots \notag \\ 
&&\hspace{120pt}\cdots\times \left( \frac{1}{2}\bar{R}^{b_{2p-2}b_{2p-1}}_{a_{2p-2}a_{2p-1}}-s^2K^{b_{2p-2}}_{a_{2p-2}}K^{b_{2p-1}}_{a_{2p-1}} \right)\,,
\end{eqnarray}
it is such that, added on top of the bulk Lagrangian, guarantees a well-posed Dirichlet principle for the boundary metric $h_{ab}$. 

Thus, the variation of Eq.(\ref{IDL}) reads
\begin{equation} \label{ILV}
\delta I=-\int\limits_{\mathcal{M}}%
d^{D}x\sqrt{-g}\,\mathcal{E}^{AB}\delta g_{AB}+%
\int\limits_{\partial\mathcal{M}}d^{d}x\,\sqrt{-h}\pi^{ab}\delta h_{ab}\,,
\end{equation}
where the equation of motion is the linear combination of the individual contributions coming from each term in the Lovelock series, that is, 
\begin{equation}\label{eom}
\mathcal{E}_{B}^{A}=\sum_{p=0}^{\floor{\frac{D-1}{2}}}\alpha_{p}\,\mathcal{E}_{(p)B}^{A}\,.
\end{equation}
In doing so, the $p$-th term in the series produces the covariantly conserved tensor
\begin{equation}\label{eomP}
\mathcal{E}_{(p)B}^{A}=-\frac{1}{2^{p+1}}%
\delta_{\lbrack BB_{1}\cdots B_{2p}]}^{[AA_{1}\cdots A_{2p}]}%
R_{A_{1}A_{2}}^{B_{1}B_{2}}\cdots R_{A_{2p-1}A_{2p}}^{B
_{2p-1}B_{2p}}\,.%
\end{equation}

The surface term in Eq.(\ref{ILV}) is proportional to the canonical momentum of the theory. This resembles standard derivations in Classical Mechanics, as the Myers term is responsible for turning the gravity action into a first-order functional \cite{GJMO}, and therefore the momentum can be readily read off as the conjugate to the metric $h_{ab}$ as 
\begin{equation} \label{canmom}
\pi^{ab}=\sum_{p=0}^{\floor{\frac{D-1}{2}}}\alpha_{p}\pi^{ab}_{(p)}\,,
\end{equation}
where, for the $p$-th Lovelock density, one obtains the associated piece for the canonical momentum
\begin{align}
\pi^{ab}_{(p)} &  =-p\int_{0}^{1}\!ds\,\delta_{[cb_{1}\cdots b_{2p-1}]}^{[aa_{1}\cdots
a_{2p-1}]}h^{cb}K_{a_{1}}^{b_{1}}\Bigg(\frac{1}{2}\bar{R}_{a_{2}a_{3}}^{b_{2}b_{3}}-s^{2} K_{a_{2}}^{b_{2}}K_{a_{3}}^{b_{3}}\Bigg)  \times
\cdots\nonumber\\
& \hspace{150pt} \cdots\times\Bigg(  \frac{1}{2}\bar{R}_{a_{2p-2}a_{2p-1}}^{b_{2p-2}%
b_{2p-1}}-s^{2} K_{a_{2p-2}}^{b_{2p-2}}K_{a_{2p-1}}^{b_{2p-1}}
\Bigg)\,.
\end{align}

The Lagrangian density associated to the action (\ref{IDL}) is rewritten employing the GCM relations and bulkanization of Myers term to obtain \cite{GJMO}
\begin{equation}
\mathcal L_{\rm 1}=\sum_{p=0}^{\floor{\frac{D-1}{2}}}\alpha_{p}\mathcal{L}^{(p)}_{\rm 1}\,,
\end{equation}
with $\mathcal{L}^{(p)}_{\rm 1}$ given by
\begin{eqnarray} \label{L_1st}
\mathcal{L}^{(p)}_{\rm 1}&=&N\sqrt{-h}\Bigg(%
\frac{1}{2^{p}}\sqrt{-h}\delta_{\lbrack b_{1}\cdots b_{2p}%
]}^{[a_{1}\cdots a_{2p}]}\bar R_{a_{1}a_{2}}^{b_{1}b_{2}}\cdots
\bar R_{a_{2p-1}a_{2p}}^{b_{2p-1}b_{2p}}+ \notag \\
&&\hspace{50pt}+2p\int_{0}^{1}\!ds\,(1-s)\delta_{\lbrack b_{1}\cdots b_{2p}]}^{[a_{1}\cdots a_{2p}]}K_{a_{1}%
}^{b_{1}}K_{a_{2}}^{b_{2}}\left(  \frac{1}{2}\bar R_{a_{3}a_{4}}^{b_{3}b_{4}}-s^{2}K_{a_{3}}^{b_{3}}K_{a_{4}}^{b_{4}}\right)\times\cdots \notag \\
&&\hspace{135pt}\cdots \times\left(  \frac
{1}{2}\bar R_{a_{2p-1}a_{2p}}^{b_{2p-1}b_{2p}}-s^{2} K_{a_{2p-1}%
}^{b_{2p-1}}K_{a_{2p}}^{b_{2p}}\right)\Bigg)\,. \label{LR}
\end{eqnarray}
Therefore, the Dirichlet action has been written explicitly as a first-order Lagrangian density. This allows to identify the tensor (\ref{canmom}) as the canonical momentum conjugate to $h_{ab}$.

The $\mathcal{E}^w_{(p)w}$ and $\mathcal{E}^w_ {(p)a}$ components of the field equations for the Gaussian frame are
\begin{eqnarray*}
\mathcal{E}^w_{(p)w} & = & -\frac{1}{2^{p+1}}\delta_{[b_{1}\cdots b_{2p}%
]}^{[a_{1}\cdots a_{2p}]}R_{a_{1}a_{2}}^{b_{1}b_{2}}\times\cdots\times
R_{a_{2p-1}a_{2p}}^{b_{2p-1}b_{2p}}\,, \\
\mathcal{E}^w_{(p)a} & = &  -\frac{1}{N}\bar{\nabla}_b \pi^b_a\,.
\end{eqnarray*}

In addition, the components $\mathcal{E}^a_{(p)b}$ are given by
\begin{eqnarray} \label{shelleomproy}
\mathcal{E}^a_{(p)b}&=&\partial_n\pi^a_{(p)b}-p\int_0^1\!ds(1-s)\delta^{[aa_1\cdots a_{2p}]}_{[bb_1\cdots b_{2p}]}%
K^{b_1}_{a_1}K^{b_2}_{a_2}\Big( \frac{1}{2}\bar R^{b_3b_4}_{a_3a_4}- s^2K^{b_3}_{a_3}%
K^{b_4}_{a_4} \Big)\times \cdots \notag \\
&& \cdots \times \Bigg( \frac{1}{2}\bar R^{b_{2p-1}b_{2p}}_{a_{2p-1}a_{2p}}-%
 s^2K^{b_{2p-1}}_{a_{2p-1}}K^{b_{2p}}_{a_{2p}}\Bigg)+\pi^b_{(p)c}%
 \left (K^{c}_{b}-K\delta^{c}_{b}\right)+\bar{\nabla}_{c}V_{b}^{ca}\,,\quad
\end{eqnarray}
where the functional $V_{b}^{ca}$ is
\begin{align}
& \qquad V_{b}^{ca}=-\frac{1}{2}\delta_{\lbrack bb_{1}\cdots b_{2p-1}%
]}^{[aca_{2}\cdots a_{2p-1}]}\bigg(K_{a_{2}}^{b_{2}}%
\bar{\nabla}^{b_{1}}K_{a_{3}}^{b_{3}}\Big(\bar{R}_{a_{4}a_{5}}^{b_{4}b_{5}%
}-2K_{a_{4}}^{b_{4}}K_{a_{5}}^{b_{5}}\Big)\times\cdots\nonumber\\
& \hspace{160pt}\cdots\times\Big(\bar{R}_{a_{2p-2}a_{2p-1}}^{b_{2p-2}b_{2p-1}%
}-2K_{a_{2p-2}}^{b_{2p-2}}K_{a_{2p-1}}^{b_{2p-1}}\Big)\bigg)\,.
\end{align}
One can also recognize in (\ref{shelleomproy}) that the second-derivative terms of the equations of motion are packed as derivatives of $\pi^{ab}$.

The addition of surface terms and its role to describe properly the physical changes across an interface is specially relevant for the next sections. The presence of a shell which is itself a boundary to the spacetime geometry will provide the link between first-order Lagrangian for gravity and the collapsing shell dynamics.

\section{Junction conditions from the variational principle} \label{sec2}
\subsection{Shell co-moving frame}

In this section, we succinctly review the kinematic description of collapsing thin shells as it appears in, e.g., Refs.\cite{Haji,RoCri}. In turn, the collapse dynamics would appear from the junction conditions expressed in terms of the discontinuity of the canonical momentum $\pi^{ab}$.

Consider a manifold $\mathcal{M}$, separated in two regions by a  spherical thin shell located at $r=R_s$. As the collapse is radial, the inner $(-)$ and outer $(+)$ regions are described by the static spherically symmetric ansatz
\begin{equation} \label{SSA}
ds^2_{\pm}=g_{\mu\nu}^{\pm}dy^\mu dy^\nu=-f^2_\pm(r)dt_{\pm}^2+\frac{dr^2}{f^2_{\pm}(r)}+r^2d\Omega_{D-2}^2\,,
\end{equation}
where $t_{\pm}$ are the corresponding time coordinates and $d\Omega^2=\omega_{mn}dx^mdx^n$ is the line element of a unit sphere $S^{D-2}$. 

On the other hand, one may take a coordinate system $x^{A}=\{\lambda,x^a\}$ of the type
\begin{equation} \label{sh}
 ds^{2}=d\lambda^{2}+\smet_{ab}(\lambda,x)\,dx^adx^b\,,
\end{equation}
adapted to the shell frame,
such that the normal direction $\lambda$ is generated by the vector $n_A=\delta^\lambda_A$. For an observer moving with the shell, the induced metric has the form
\begin{equation} \label{sha}
 ds^{2}\big |_{\lambda=0}=\smet_{ab}dx^adx^b=-d\tau^2+R_s^2d\Omega_{D-2}^2\,,
\end{equation}
with $x^a=\{\tau,\theta_{1},...,\theta_{D-2}\}$. 

One may parametrize the shell position using its proper time $\tau$, in terms of the Schwarzschild-like coordinates $r=R_s(\tau)$ and $t_{\pm}=t_{\pm}(\tau)$. 
Indeed, gluing the line elements Eq.(\ref{sh}) and Eq.(\ref{SSA}), one obtains the relation
\begin{equation} \label{shellcon}
f^2_{\pm}(R_s)\dot{t}_{\pm}^2-\frac{\dot{R}_s^2}{f^2_{\pm}(R_s)}=1\,,
\end{equation}
where the dot stands for $\partial_\tau$. 
In the following analysis, we drop the subscript $\pm$ in the metric function, as the treatment applies equally to either the interior or exterior regions. 

Of particular usefulness is the definition of the factor
\begin{equation} \label{gammapm}
\gamma=f^2(R_s)\dot{t}=\sqrt{\dot{R_s}^2+f^2(R_s)}\,,
\end{equation}
which reduces to the standard relativistic factor when the spacetime is Minkowski.
The normal to the shell is a space-like unit vector which, as described in the coordinate system of the static black hole geometry $n_\mu=\frac{\partial x^A}{\partial y^\mu} n_A$, takes the form
\begin{equation}
 n_\mu=\left (\frac{\gamma}{f^2(R_s)},-\dot{R_s},\vec{0}\right )\,.
\end{equation}

By definition, the extrinsic curvature of the shell geometry is given by
\begin{equation} \label{KabE}
K_{ab}=e_{a}^{\mu}e_{b}^{\nu}\nabla_{\mu}n_{\nu}\,,
\end{equation}
in terms of the local orthonormal basis $e_{a}^{\mu}$
\begin{equation}
e_a^\mu=\frac{\partial y^\mu}{\partial x^a}\,,
\end{equation}
which projects the spacetime indices to the ones of the shell\footnote{See Appendix \ref{Cfc} for a detailed mapping between these two coordinate frames}.

The non-vanishing components of the extrinsic curvature, computed from Eq.(\ref{KabE}), take the form 
\begin{align}
K^\tau_{\tau}  &  =-\gamma^{\prime}\,,\\
K^n_{m}  &  =-\gamma R_s\delta^n_{m}\,.
\end{align}
In addition, the components of the curvature tensor for the intrinsic geometry of the shell are
\begin{equation}
\bar{R}_{\tau n}^{\tau m}=\frac{\ddot{R_s}}{R_s}\delta_{n}^{m}\quad
,\quad\bar{R}_{pq}^{nm}=\frac{1+\dot{R_s}^{2}}{R_s^{2}}\delta_{\lbrack
pq]}^{[nm]}\,.
\end{equation}
The energy-momentum tensor sourcing the discontinuity of the geometry is given by
\begin{equation}
T^{AB}=\frac{2}{\sqrt{-g}}\frac{\delta \mathcal{L}_{M}}{\delta g_{AB}}=S^{AB} \delta(\lambda)\,,
\end{equation}
where 
$\mathcal{L}_M$ is a generic matter Lagrangian density. The matter distribution is such that it appears localized on the shell. 


The interface that divides the spacetime is given by a physical thin shell, whose stress tensor is the one of a perfect fluid
\begin{equation}
S^{ab}=(\sigma+p)u^au^b+p\,h^{ab}\,, \label{emta}%
\end{equation}
for a velocity vector $u^a$.
The conservation of this energy-momentum tensor, written in the coordinate frame $\{x^a\}$ $(\nabla_a T^{ab}=0)$, leads to the continuity equation
\begin{equation}
    \nabla_a \big(\sigma u^a\big)-\nabla_a \big(p u^a\big)=0\,.
\end{equation}
In the co-moving frame, where the observer is at rest respect to the shell the relevant component of the stress tensor is

\begin{equation}
T_{\tau}^{\tau}=-\sigma\delta(\lambda)\,. \label{emt}%
\end{equation}
This relation, combined with the corresponding equation of state for the shell matter will determine the collapse dynamics, once the junction conditions are imposed.

\subsection{Discontinuity of the canonical momentum}

As rendered manifest by the previous discussion, the use of Schwarzschild-like coordinates is suitable to describe both the interior and exterior regions in the present collapse setup. Of course, this metric can be put in a Gauss-normal form for a radial foliation of the spacetime.
In turn, in the shell co-moving frame, the normal direction to the shell is $\lambda$ which mixes up the radial and time Schwarzchild coordinates. Thus, junction conditions will be derived from a variational principle in adapted Gauss normal frame. 

For any Lovelock theory written in first-order form (\ref{L_1st}) and taking the shell as a boundary, the variation of the action is given by 
\footnote{At radial infinity, yet another Gibbons-Hawking-Myers term is needed (in Schwarzchild-like radial coordinates) to ensure a well defined variational problem for Dirichlet boundary conditions on $h_{ij}$.}

\begin{equation} \label{varprin}
\delta I=\frac{1}{16 \pi G}\int\limits_{\mathcal{M}}d^Dx\,\partial_\lambda\big(\sqrt{-h}\pi^{ab}\delta h_{ab}\big)\,,
\end{equation}
provided that the equations of motion hold. 
The integration in an infinitesimal interval $(-\varepsilon,+\varepsilon)$ across the normal direction yields
\begin{equation} \label{alm}
\delta I=\frac{1}{16 \pi G}\int\limits_{\shell}d^dx\,\sqrt{-\smet}\bigg(\pi^{ab}_+\delta \smet^+_{ab}-\pi^{ab}_-\delta \smet^-_{ab}\bigg)\,,
\end{equation}
where one defines the limit by the left and by the right of $\smet_{ab}$ as
\begin{align}
\smet_{ab}^{+}  =\lim_{\lambda\rightarrow
0}\smet_{ab}^{+}\left( \lambda,x^a\right)
\nonumber \quad , \quad
\smet_{ab}^{-} =\lim_{\lambda\rightarrow
0}\smet_{ab}^{-}\left(  -\lambda,x^a\right)\,.
\end{align}

Then, the (Riemannian) condition of $g$ being smooth is relaxed and one considers instead a metric which is continuous at the shell location, that is,
\begin{equation}
\smet_{ab}^{+}= \smet_{ab}^{-}=\smet_{ab}\,,
\end{equation}

Localizing the energy-momentum tensor with a delta function at the shell, as given by Eq. (\ref{emt}), renders Eq.(\ref{alm}) into the following form
\begin{equation}
\delta I +\delta I_{M}=\frac{1}{16 \pi G}\int\limits_{\shell}d^dy\,\sqrt{-\smet}\Big(\jump{\pi^{ab}}-8\pi G S^{ab}\,\Big)\delta \smet_{ab}\,.
\end{equation}
Since variations of the metric $\delta \smet_{ab}$ are arbitrary, the variational principle requires  
\begin{equation}\label{junccon}
\jump{\pi^{ab}}=8\pi G S^{ab}\,.
\end{equation}


In this way, the condition $\delta I+\delta I_{M}=0$ is met by imposing the junction conditions at $\lambda=\lambda(R_s)$, and the standard  Dirichlet condition on the metric at radial infinity. In turn, the junction condition (\ref{junccon}) describes the motion of the collapsing shell hypersurface. 

\section{Junction conditions in Lovelock gravity} \label{sec3}

\subsection{General Relativity}
As a warm up exercise, one may consider the gravitational collapse of thin shells in Einstein theory of gravity with negative cosmological constant.

As the matter density is given by a delta function the geometry \ref{SSA} will jump across the shell characterized by the discontinuity in the metric function
\begin{equation}
f_\pm^2(r)=1+\frac{r^2}{\ell^2}-\frac{16\pi G M_\pm}{(D-2)\Omega_{D-2} r^{D-3}}\,,
\end{equation}
where $\Omega_{D-2}$ is the volume of the unit sphere $S^{D-2}$. 
The difference in the mass between the exterior and interior black holes is sourced by the $(\tau,\tau)$ component of the stress tensor, what leads to 
\begin{equation}
\pi_{\tau}^{\tau}=(D-2)\frac{\gamma}{R_{s}}\,.  
\end{equation}
Thus, the junction condition (\ref{junccon}) for General Relativity reduces to the relation 
\begin{equation}
\jump{\pi_{\tau}^{\tau}}  =  (D-2)  \frac{1}{R_{s}}\left(\gamma_+-\gamma_- \right) \,=-8\pi G\sigma 
\end{equation}
i.e., 
\begin{equation}
\gamma_+-\gamma_-=-\frac{8\pi G\sigma}{D-2}R_s\,,
\end{equation}
in agreement with Refs.\cite{Isra,RoCri,Lin:2010wy}.

One can work out an alternative form, multiplying by $\gamma_++\gamma_-$. This is particularly convenient in the case of incoherent dust (no pressure). Indeed, in that situation, one obtains the difference between the inner and outer mass
\begin{eqnarray}
\Delta M = M_{+}-M_{-} = m\frac{\big(\gamma_++\gamma_-\big)}{2}\,,
\end{eqnarray}
in terms of the proper mass of the shell, $m=R_s^{D-2}\Omega_{D-2} \sigma$.

\subsection{The Einstein-Gauss-Bonnet gravity} \label{EGBJC}

In the Lovelock series, the next gravity theory to be considered is Gauss-Bonnet. As one is interested in this term as a correction to GR appearing in the bulk action, its addition define Einstein-Gauss-Bonnet gravity. For the collapse of thin shells with spherical symmetry, the solution with the same symmetry is given by Boulware-Deser one  \cite{BoulDer} 
\begin{eqnarray}
f^2_\pm(r)&=&1+\frac{r^2}{2\tilde \alpha}\Bigg (1+ \sigma \sqrt{1-4\tilde\alpha\left(\frac{1}{\ell^2}-\frac{16\pi G M_\pm}{(D-2)\Omega_{D-2} r^{D-1}}\right )}\Bigg )\,.
\end{eqnarray}
where $\tilde \alpha=\alpha(D-3)(D-4)$ and $\sigma=\pm 1$, which represents two branches of the theory. For given values of the GB coupling $\alpha$ and the cosmological constant $\Lambda$, the theory may feature black hole solutions.
As for the present treatment, one may assume the existence of such black holes. The results then are is given in terms of the function $\gamma$, defined in Eq.(\ref{gammapm}), and they equally apply for any function $f^2(r)$.

The component $(\tau,\tau)$ of the canonical momentum has the form
\begin{equation}
\pi_{\tau}^{\tau} =(D-2)\frac{\gamma}{R_{s}}+2\tilde\alpha (D-2)\frac{\gamma}{R_{s}^3}\left(  1+\dot{R}_{s}%
^{2}-\frac{1}{3}\gamma^{2}\right)  \,,\notag \\
\end{equation}
such that the junction condition can then be written as
\begin{align} \label{EGB_junccon}
-8\pi G\sigma & =(\gamma_{+}-\gamma_{-})(D-2)\Bigg (  R_{s}^{-1}+  2\tilde\alpha R_{s}^{-3}\left(  1+\dot{R}_{s}^{2}-\frac
{1}{3}\left(  \gamma_{+}^{2}+\gamma_{+}\gamma_{-}+\gamma_{-}^{2}\right)
\right)  \Bigg )  \,.
\end{align}
This expression governs the shell dynamics in EGB gravity, and it consistently reproduces the results in the existing literature \cite{GaGiGraWi,Lin:2010wy}.
In the case the theory possesses a unique vacuum, which corresponds to the particular value for the GB coupling $\tilde\alpha=\frac{\ell^2}{4}$, the above expression reduces to the relation
\begin{equation}
-8\pi G\sigma=(\gamma_{+}-\gamma_{-})(D-2)\left(  R_{s}^{-1}+\frac{\ell
^{2}R_{s}^{-3}}{2}\left(  1+\dot{R}_{s}^{2}-\frac{1}{3}\left(  \gamma_{+}%
^{2}+\gamma_{+}\gamma_{-}+\gamma_{-}^{2}\right)  \right)  \right)  \,.
\end{equation}
The metric function for the black solutions around this critical point in the parametric space features a distinctive asymptotic behavior in the mass term, with a much slower falloff for large $r$. Thus, even though it is neither possible to solve explicitly $\dot{R}_{s}^{2}$ nor to find a closed expression for $\Delta M$, it is expected that the dynamics will be radically different for that case.

\subsection{General Lovelock case}

For a generic Lovelock theory of gravity, the canonical momentum is the linear combination of the corresponding term associated to every $p-$th term of the series
\begin{equation}
\pi_{\tau}^{\tau}=\sum_{p=0}^{\floor{\frac{D-1}{2}}}\pi_{(p)\tau}^{\tau}\,, \label{pitautauLove}
\end{equation}
with the contribution due to the $p-$th term in the Lovelock Lagrangian given by
\begin{equation}\label{pilov}
\pi_{(p)\tau}^{\tau}  =\frac{p\alpha_{p}(D-2)!}{(D-2p-1)!}\frac{\gamma} {R_{s}^{2p-1}}\int_{0}^{1}dt\,\left(  1+\dot{R}_{s}^{2}-\gamma^{2}%
t^{2}\right) ^{p-1}\,.
\end{equation}
As a consequence, the junction condition for Lovelock gravity is expressed in terms of the following discontinuity
\begin{equation} \label{Lovelock_junctCond}
  \sum_{p=0}^{\floor{\frac{D-1}{2}}}\frac{p\alpha_{p}(D-2)!}{(D-2p-1)!}\jump{\frac{\gamma}{R_{s}^{2p-1}}\int_{0}^{1}dt\,\left(  1+\dot{R}_{s}^{2}-\gamma^{2}t^{2}\right)
^{p-1}}  =-8\pi G\sigma \,.
\end{equation}

A particular choice of the set of couplings in the Lovelock series leads to the so-called Lovelock Unique Vacum (LUV) theory \cite{BHS}. This choice intends to free higher-curvature gravity from undesirable instabilities that may trigger transitions between different vacua. The price to pay is that now global AdS space is a zero of the field equations with multiplicity $k$. Therefore, this class of gravity theories do not accept a linearization around AdS background, such that their black holes do not have the asymptotic behavior of the Schwarzschild solution.

As for the momentum tensor, the generic formula (\ref{pitautauLove}) turns into
\begin{eqnarray}
\pi^\tau_{\tau}&=&\frac{(D-2)}{k\ell^2}\sum_{p=0}^kp\,\ell^{2p}%
{{k}\choose{p}}\frac{\gamma}{R_s^{2p-1}}\int_0^1dt\,%
\left (1+\dot{R}_s^2-\gamma^2t^2\right )^{p-1} \,.\notag
\end{eqnarray}
In this case, the sum can be factorized using the binomial expansion, such that
the junction condition (\ref{junccon}) for LUV theories can be cast in the form 
\begin{equation} \label{LUV_junctCond}
(D-2)\left(  \frac{\ell^{2}}{R_{s}^{2}}\right)  ^{k-1}\jump{\frac{\gamma}{
R_{s}}\int_{0}^{1}dt\,\left(  1+\frac{R_{s}^{2}}{\ell^{2}%
}+\dot{R}_{s}^{2}-\gamma^{2}t^{2}\right)  ^{k-1}}
=-8\pi G\sigma
\end{equation}

As the relativistic factor $\gamma
_{+}$ and $\gamma_{-}$ depend on $\dot{R}_{s}^{2}\,$, $M_{+}$ and $M_{-}$, it turns a difficult task to properly isolate the shell velocity and, therefore, to determine the exact collapse dynamics from the initial conditions. 

\subsection{Comparison to existing literature}

In the ref.\cite{CSD} a different path to the derivation of the junction conditions in Lovelock gravity was taken. The procedure followed by these authors considers --as a starting point-- the equations of motion (\ref{eomP}), written down in the static black hole ansatz (\ref{SSA})

\begin{eqnarray} \label{new1}
\mathcal{E}^t_{(p)t}&= &
\partial_n\big(\pi^t_{(p)t}\big) +(D-2)\frac{f(r)\pi^t_{(p)t}}{r}+\hspace{120pt} \notag \\%
&&\hspace{60pt}-p\frac{(D-2)!}{(D-2p-2)!}\frac{f^2(r)}{r^{2p}}%
\int_0^1ds\,(1-s)\,\left (1-s^2f^2(r)\right )^{p-1}\,,
\end{eqnarray}
in Schwarzschild-like coordinates $(t,r)$, where the canonical momentum is given by
\begin{equation*}
\pi_{(p)t}^{t}=p\int_{0}^{1}ds\frac{(D-2)!}{(D-2p-1)!}\frac{f(r)}{r^{2p-1}}\left(  1-s^{2}f^{2}(r)\right)^{p-1}\,.
\end{equation*}

The point is that, for the specific ansatz taken, the component $(t,t)$ of the field equations can be further simplified as single derivative term

\begin{equation} \label{Cris_etal}
\mathcal{E}^t_{t}= -\frac{(D-2)!}{2r^{D-2}} \frac{d}{dr} \left( \sum_{p=0}^{\floor{\frac{D-1}{2}}}\frac{\alpha_p}{(D-2p-1)!} \left( r^{D-2p-1}( 1-f^2(r))^{p} \right) \right) \,.
\end{equation}
Once this total derivative was obtained, the idea in Ref.\cite{CSD} was to perform a subsequent integration across the shell position, along the normal direction $\lambda$. This integration requires a projection between the coordinate systems $(t,r)$ and $(\tau,\lambda)$ (a sort of \emph{rotation}) which gives rise to additional factors depending on $\gamma$. Because, in this ansatz, the terms on top of $\partial_n\pi^t_{(p)t}$ in Eq.(\ref{new1}) accidentally contribute to a total derivative, the matching conditions for the shell in Ref.\cite{CSD} are mistakenly proportional to $\Delta M$. Conversely, the canonical momentum cannot account --only by itself-- for the difference in mass between the interior and the exterior regions, as the energy of the system in Lovelock gravity strongly depends on the multiplicity/degeneracy of the vacuum state  \cite{Arenas2017,Arenas2019}.

An unambiguous way to rederive the junction conditions would require writing down the corresponding field equation in an schematic form
\begin{equation}
\mathcal{E}_{t}^{t}=\sum_{p=0}^{\floor{\frac{D-1}{2}}} \mathcal{E}_{(p)t}^{t}=\partial_n\left(  \pi_{t}^{t}\right)+\text{bounded terms}=\text{delta
source}\,.
\label{accidental_terms}
\end{equation}
Here, the \emph{bounded terms} are contributions which go to zero after one performs the integration across the shell and takes the zero thickness limit, regardless the particular ansatz used. As a matter of fact, the above relation is an efficient form to pack the second-derivative terms (in the radial direction) as normal derivatives of the momentum. This is consistent with the picture developed by Deruelle et al. in Ref.\cite{Deruelle:2007pt}, where junction conditions in higher-derivative gravity can be readily obtained from the highest-derivate term.

\section{Conclusions}

In this work, we have exploited the connection between junction conditions and the variational principle for a Dirichlet boundary condition on the metric. As a direct consequence of this treatment, we have studied the thin shell collapse in Lovelock gravity as coming from the discontinuity in the canonical momentum of the theory.

After reviewing, from the above standpoint, the derivation of shell dynamics  in General Relativity and Einstein-Gauss-Bonnet gravity, we work out the corresponding expression for an arbitrary Lovelock theory. These derivations stress the link between the  junction conditions and the discontinuity of the canonical momentum, which is properly identified once the Lagrangian adopts a first-order form. 

In General Relativity, a spherically symmetry ansatz leads necessarily to the formation of a Schwarzschild solution, as dictated by the Birkhoff theorem. The only parameter accounting for global properties of the geometry is the black hole mass. Junction conditions in Einstein gravity sees the jump between the inner and outer black hole mass as proportional to the mass of the shell \cite{Isra,RoCri}. Later work derives a similar relation for LUV gravity theories \cite{CSD}.
In bold contrast to these results, the shell dynamics developed here involves nonlinear combinations of the $\gamma$ factors.
Therefore, the erroneous reasoning in ref.\cite{CSD} is assuming that junction conditions can be readily read off from the total derivative term in the field equations for the static, spherically symmetric ansatz. Because the total derivative term acquires extra, accidental contributions on top of $\pi^{ab}$, the picture of the junctions conditions as associated to the canonical momentum is lost.

As a prospect, it would be interesting to explore the physical implications of gravitational collapse of thin shells within a holographic framework, equipped with the tools presented here. In Ref.\cite{Holoshells}, the authors study thermalization in a boundary Conformal Field Theory, where the bulk spacetime is a solution to AdS gravity. In particular, this provides a dual gravitational setup in order to work out the time evolution of entanglement entropy of the boundary CFT. It is indeed a quite appealing idea to think of  a similar model where higher curvature terms are included in the bulk gravity action \cite{AAMO}.  As discussed in the previous section, the correct identification of the discontinuous quantities in the bulk geometry  may turn of key importance when it comes to a proper holographic description of the system. 

\section*{Acknowledgements}

The authors would like to thank Cristóbal Corral, Nathalie Deruelle and Olivera Miskovic for discussions and for insightful comments.
P.G. was funded by UNAB Ph.D. Scholarship 2018-2020 and ANID Ph.D. Scholarship 21211807.
N.M. was funded by the Fondecyt Iniciación Grant 11180894 \textit{Covariant boundary terms and conservation laws in modified gravity}. R.O. was funded in part by ANID Regular Grant 1090533 \textit{Black holes and asymptotic symmetries} and Anillo ANID-SCIA-ACT210100 \textit{Holography and its applications to High Energy Physics, Quantum Gravity and Condensed Matter Systems.}

\section*{Appendices}
\appendix

\section{Kronecker delta of rank $p$} \label{DK}

The totally-antisymmetric Kronecker delta of rank $p$ is defined as the
determinant
\begin{equation}
\delta _{\left[A _{1}\cdots A _{p}\right] }^{\left[ B _{1}\cdots B
_{p}\right] }:=\left\vert
\begin{array}{cccc}
\delta _{A _{1}}^{B _{1}} & \delta _{A _{1}}^{B _{2}} & \cdots &
\delta _{A _{1}}^{B _{p}} \\
\delta _{A _{2}}^{B _{1}} & \delta _{A _{2}}^{B _{2}} &  & \delta
_{A _{2}}^{B _{p}} \\
\vdots &  & \ddots &  \\
\delta _{A _{p}}^{B _{1}} & \delta _{A _{p}}^{B _{2}} & \cdots &
\delta _{A _{p}}^{B _{p}}%
\end{array}%
\right\vert \,.
\end{equation}%
A contraction of $k\leq p$ indices in the Kronecker delta of rank $p$
produces a delta of rank $p-k$,
\begin{equation}
\delta _{\left[ A _{1}\cdots A _{k}\cdots A _{p}\right] }^{\left[B
_{1}\cdots B _{k}\cdots B _{p}\right] }\,\delta _{B _{1}}^{A
_{1}}\cdots \delta _{B _{k}}^{A _{k}}=\frac{\left( N-p+k\right) !}{%
\left( N-p\right) !}\,\delta _{\left[ A _{k+1}\cdots A _{p}\right] }^{%
\left[ B _{k+1}\cdots B _{p}\right] }\,,
\end{equation}%
where $N$ is the range of indices.

\section{Gauss-normal coordinates and  Gauss-Codazzi-Mainardi relations} \label{GC}
For the spacelike foliation (\ref{GN}),
the components of the Christoffel symbol are
\begin{eqnarray}
\Gamma^w_{ww} = \frac{\partial_wN}{N}\,, & \quad & \Gamma^w_{ab}=\frac{1}{N}K_{ab}\, \notag \\
\Gamma^a_{bw} = -NK^a_b\,, & \quad & \Gamma^a_{bk}=\Gamma^a_{bk}(h)\,.
\end{eqnarray}
where $K_{ab}$ is the extrinsic curvature
\begin{equation} \label{extrin}
K_{ab}=-\frac{1}{2N}\partial_w h_{ab}\,.
\end{equation}

The curvature tensors are given by
\begin{eqnarray} \label{riemann}
R^{wa}_{wb} = \frac{1}{N}\left ( K^a_b\right )'-K^a_cK^c_b\,, & \quad &
R^{wa}_{bc} = \frac{1}{N}\left ( \bar{\nabla}_b K^a_c-\bar{\nabla}_c K^a_b\right )\,, \notag \\
R^{ab}_{wc} = N\left ( \bar{\nabla}^a K^b_c-\bar{\nabla}^b K^a_c\right )\,, & \quad &
R^{ab}_{cd}= \bar{R}^{ab}_{cd}(h)-K^a_cK^b_d+K^a_dK^b_c\,.
\end{eqnarray}
In the notation used here, the prime stands for a partial derivative in $w$, $\bar{\nabla}$ is the covariant derivative defined with the connection $\Gamma^a_{bc}(h)$ associated to the boundary metric,
and $\bar{R}_{abcd}$ is the boundary Riemann tensor. Boundary indices are raised or lowered with the metric $h_{ab}$.
Then, the spacetime Ricci scalar and Ricci tensor can be expressed as
\begin{eqnarray} \label{ricci}
R^a_b &=& \bar{R}^a_b(h)-K^a_bK+\frac{1}{N}\partial_w(K^a_b)\,, \notag \\
R^w_a &=& \frac{1}{N}\left (\bar{\nabla}_a K - \bar{\nabla}_c K^c_a\right )\,, \notag \\
R^w_w &=& \frac{1}{N}K'-K^c_dK^d_c\,, \notag \\
R &=& \bar{R}(h)-K^2-K^c_dK^d_c+\frac{2}{N}\partial_wK\,.
\end{eqnarray}

In turn, for the static black hole ansatz (\ref{SSA}), the nonvanishing components of the Christoffel symbol, in Schwarzschild coordinates, are
\begin{eqnarray}
\Gamma^r_{rr}=\frac{-\partial_rf(r)}{f(r)} & \quad & \Gamma^t_{rt}=\frac{\partial_rf(r)}{f(r)} \\
\Gamma^r_{tt}=f^3(r)\partial_rf(r) & \quad & %
\Gamma^n_{rm}=\frac{1}{r}\delta^n_m \\
\Gamma^r_{nm}=-f^2(r)r\,\omega_{nm} & \quad & \Gamma^p_{nm}=\Gamma^p_{nm}(\omega)
\end{eqnarray}

\section{Co-moving frame coordinates and continuity conditions} \label{Cfc}

One can always introduce a set of adapted Gaussian coordinates $x^{A}=\{\lambda,x^a\}=\{\lambda,\tau,\theta^{(m)}\}$ and refer them to the Schwarzschild-like coordinates of the static interior/exterior geometry, as discussed in Ref.\cite{Holoshells}. This is done by considering the change of coordinates in the bulk
\begin{equation}
r=r\left(  \lambda,\tau\right)  \text{ and }t=t\left(  \lambda,\tau\right)\,.
\label{rttaulam}%
\end{equation}
In doing so, the metric (\ref{SSA}) in the proper coordinates is given by
\begin{eqnarray} \label{gAB}
ds^2={g}_{AB}dx^Adx^B&=&-\left(f^2(r)\dot{t}^2-\frac{\dot{r}^2}{f^2(r)}\right )d\tau^2+
2\left (\frac{\dot{r}\partial_\lambda r}{f^2(r)}-f^2(r)\dot{t}%
\partial_\lambda t\right )d\tau d\lambda \notag \\
& & \hspace{50pt} +\left (\frac{\big(\partial_\lambda r\big)^2}{f^2(r)}-f^2(r)%
\big(\partial_\lambda t\big)^2\right )d\lambda^2+r^2d\Omega^2\,.
\end{eqnarray}
Matching the above line element to the one in the adapted coordinates
\begin{equation} \label{g_adapted}
ds^{2}=d\lambda^{2}+\smet_{ab}\,dx^adx^b\,=d\lambda^{2}-d\tau^{2}+R_s^{2}\left(\lambda,\tau\right)d\Omega^{2}\,,%
\end{equation}
leads to the conditions
\begin{align} \label{yAc}
f^{2}\dot{t}^{2}-\frac{\dot{R_s}^{2}}{f^{2}}  =1\quad ,\quad
\frac{\dot{R_s}\partial_{\lambda}R_s}{f^{2}}-f^{2}\dot{t}\partial_{\lambda}t 
=0 \quad , \quad
\frac{\left(  \partial_{\lambda}R_s\right)  ^{2}}{f^{2}}-f^{2}\left(
\partial_{\lambda}t\right)  ^{2}  =1\,.
\end{align}

It is straightforward to check that this treatment recovers the shell metric Eq.(\ref{sh}) for a constant $\lambda$. Therefore, without loss of generality, the shell is located at $\lambda=0$.


\begin{thebibliography}{99}
\bibitem{Oppen} J.R. Oppenheimer and H. Snyder, \textit{On Continued gravitational contraction}, Phys. Rev. \textbf{56}, 455 (1939).

\bibitem{Isra} W. Israel, \textit{Singular hypersurfaces and thin shells in general relativity}, Nuovo Cim. \textbf{B44}, 1 (1966);
W. Israel, \textit{Gravitational Collapse and Causality}, Phys. Rev. \textbf{153}, 1388 (1967). 
\bibitem{Shiro} T. Shiromizu, K. Maeda and M. Sasaki, \textit{The Einstein equations on the 3-brane world}, Phys. Rev \textbf{D62}, 024012 (2000). [gr-qc/9910076]

\bibitem{Parry:2005eb} M.~Parry, S.~Pichler and D.~Deeg, \textit{Higher-derivative gravity in brane world models}, JCAP \textbf{04}, 014 (2005). [hep-ph/0502048]
\bibitem{Nojiri:2001ae} S.~Nojiri, S.~D.~Odintsov and S.~Ogushi,
 \textit{Cosmological and black hole brane world universes in higher derivative gravity}, Phys. Rev. \textbf{D65}, 023521 (2002). 	[hep-th/0108172]

\bibitem{Deruelle:2007pt} N.~Deruelle, M.~Sasaki and Y.~Sendouda,
\textit{Junction conditions in f(R) theories of gravity}, Prog. Theor. Phys. \textbf{119}, 237 (2008). [arXiv:0711.1150]
\bibitem{LL} D. Lovelock, \textit{The Einstein tensor and its generalizations}, J. Math. Phys. \textbf{12}, 498 (1971).

\bibitem{Omar} O.~Rodr\'iguez-Tzompantzi, \textit{Conserved laws and dynamical structure of axions coupled to photons}, Int. J. Mod. Phys. \textbf{A36}, 2150259 (2021). [arXiv:2001.07101]

\bibitem{Myers} R. Myers, \textit{Higher-derivative gravity, surface terms, and string theory}, Phys. Rev. \textbf{D36}, 392 (1987).

\bibitem{Davis} S. C. Davis, \textit{Generalised Israel Junction Conditions for a Gauss-Bonnet Brane World}, Phys. Rev. \textbf{D67}, 024030 (2003). [hep-th/0208205]

\bibitem{Charmousis:2002rc} C.~Charmousis and J.~F.~Dufaux, \textit{General Gauss-Bonnet brane cosmology},
Class. Quant. Grav. \textbf{19}, 4671 (2002).

\bibitem{Gravanis:2002wy} E.~Gravanis and S.~Willison, \textit{Israel conditions for the Gauss-Bonnet theory and the Friedmann equation on the brane universe}, Phys. Lett. B \textbf{562}, 118 (2003). 	[hep-th/0209076]

\bibitem{GaGiGraWi} C. Garraffo, G. Giribet, E. Gravanis and S. Willison, \textit{Vacuum thin shell solutions in five-dimensional Lovelock gravity}. [arXiv:1001.3096]

\bibitem{Lin:2010wy} F. Lin, C. Wang and C. Yeh, \textit{Breathing Vacuum Bubbles in Five-Dimensional Gauss-Bonnet Gravity}. [arXiv:1003.4402]

\bibitem{MuellerHoissen:1989yv}
F.~Mueller-Hoissen, \textit{Gravity Actions, Boundary Terms and Second Order Field Equations}, Nucl. Phys. \textbf{B337}, 709 (1990).

\bibitem{OlRo} O. Miskovic and R. Olea, \textit{Counterterms in dimensionally continued AdS gravity}, JHEP \textbf{10}, 028 (2007). 	[arXiv:0706.4460]

\bibitem{TeiZan} C. Teitelboim and J. Zanelli, \textit{Dimensionally continued topological gravitation theory in Hamiltonian form}, Class. Quant. Grav. \textbf{4}, L125 (1987). 

\bibitem{BroYo} J.D. Brown and J.W. York, Jr.,
\textit{Quasilocal energy and conserved charges derived from the gravitational action}, Phys. Rev. \textbf{D47}, 1407 (1993). [gr-qc/9209012]

\bibitem{AMOP} G.~Anastasiou, O.~Miskovic, R.~Olea and I.~Papadimitriou, \textit{Counterterms, Kounterterms, and the variational problem in AdS gravity}, JHEP \textbf{08}, 061 (2020). [arXiv:2003.06425]

\bibitem{LiuSab}{J. Liu and W. Sabra, \textit{Hamilton–Jacobi counterterms for Einstein–Gauss–Bonnet gravity}, Class. Quantum Grav. \textbf{27}, 175014 (2010). [arXiv:0807.1256]}

\bibitem{Padmanabhan:2010zzb} T.~Padmanabhan, \textit{Gravitation: Foundations and frontiers}, Published in: Cambridge, UK: Cambridge Univ. Pr. (2010) 700p, ISBN: 9787301227879.

\bibitem{GJMO} P.~Guilleminot, F.~L.~Juli\'e, N.~Merino and R.~Olea, \textit{First-order Lagrangian and Hamiltonian of Lovelock gravity}, Class. Quant. Grav. \textbf{38}, 10 (2021). [arXiv:2011.01296]

\bibitem{York:1972sj} J.~W.~York, Jr., \textit{Role of conformal three geometry in the dynamics of gravitation}, Phys. Rev. Lett. \textbf{28}, 1082 (1972).

\bibitem{Gibbons:1976ue} G.~W.~Gibbons and S.~W.~Hawking, \textit{Action Integrals and Partition Functions in Quantum Gravity}, Phys. Rev. \textbf{D15}, 2752 (1977).

\bibitem{ADM} R. Arnowitt, S. Deser, and C. W. Misner, \textit{Dynamical Structure and Definition of Energy in General Relativity}, Phys. Rev. \textbf{116}, 1322 (1959).

\bibitem{Haji} P. H\'aj\'i\v{c}ek and J. Bi\v{c}\'ak, \textit{Gauge Invariant Hamiltonian Formalism for Spherically Symmetric Gravitating Shells}, Phys. Rev. \textbf{D56}, 4706 (1997). [gr-qc/9706022]

\bibitem{BoulDer}  D. Boulware and S. Deser, \textit{String Generated Gravity Models}, Phys. Rev. Lett. \textbf{55}, 2656 (1985).

\bibitem{BHS} J. Cris\'ostomo, R. Troncoso, and J. Zanelli, \textit{Black hole scan}, Phys. Rev. \textbf{D62}, 084013 (2000). [hep-th/0003271]

\bibitem{RoCri} J. Cris\'ostomo and R. Olea, \textit{Hamiltonian Treatment of the Gravitational Collapse of Thin Shells}, Phys. Rev. \textbf{D69}, 104023 (2004). [hep-th/0311054]

\bibitem{CSD} J. Cris\'ostomo, S. del Campo and J. Saavedra, \textit{Hamiltonian treatment of collapsing thin shells in Lanczos-Lovelock theories}, Phys. Rev. \textbf{D70}, 064034 (2004). [hep-th/0311259]

\bibitem{BriRa} Y. Brihaye and E. Radu,\textit{Black objects in the Einstein-Gauss-Bonnet theory with negative cosmological constant and the boundary counterterm method}, JHEP \textbf{09}, 6 (2008). 	[arXiv:0806.1396]

\bibitem{Holoshells} V. Keranen et al., \textit{Gravitational collapse of thin shells: Time evolution of the holographic entanglement entropy}, JHEP \textbf{1506}, 126 (2015). [arXiv:1502.01277]

\bibitem{bh01} C. Callan, I. Klebanov and M. Perry, \textit{String Theory Effective Actions}, Nucl. Phys. \textbf{B278}, 78 (1986).

\bibitem{bh02} P. Candelas, G. Horowitz, A. Strominger and E. Witten, \textit{Vacuum Configurations for
Superstrings}, Nucl. Phys. \textbf{B258}, 46 (1985).

\bibitem{bh03}  D. Gross and J. Sloan, \textit{The Quartic Effective Action for the Heterotic String}, Nucl. Phys.
textbf{B291}, 41 (1987).

\bibitem{bh04} M. Guica, L. Huang, W. Li and A. Strominger, \textit{Rˆ2 Corrections for 5D Black Holes and
Rings}, JHEP \textbf{0610}, 036 (2006). [hep-th/0505188]

\bibitem{bh05} D. Wiltshire, \textit{Spherically Symmetric Solutions Of Einstein-Maxwell Theory With A
Gauss-Bonnet Term}, Phys. Lett. \textbf{B169}, 36 (1986).

\bibitem{bh08} M. Aiello, R. Ferraro and G. Giribet, \textit{Exact Solutions of Lovelock-Born-Infeld Black Holes},
Phys. Rev. \textbf{D70}, 104014 (2004). [gr-qc/0408078]

\bibitem{bh07} Rong-Gen Cai, \textit{Gauss-Bonnet Black Holes in AdS Spaces}, Phys. Rev. \textbf{D65}, 084014 (2002). [hep-th/0109133]

\bibitem{bh09} Rong-Gen Cai and N. Ohta, \textit{Black holes in pure Lovelock gravities}, Phys. Rev. \textbf{D74}, 064001 (2006). 	[hep-th/0604088]

\bibitem{bh10} M. Bañados, C. Teitelboim and J. Zanelli, \textit{Dimensionally continued black holes}, Phys. Rev. \textit{D49}, 975 (1994). [gr-qc/9307033]

\bibitem{AAMO} G. Anastasiou, I.J. Araya, R.B. Mann and R. Olea, \textit{Renormalized holographic entanglement entropy in Lovelock gravity}, JHEP \textbf{2106}, 073 (2021). [arXiv:2103.14640]

\bibitem{Arenas2017}
G.~Arenas-Henriquez, O.~Miskovic and R.~Olea,
\textit{Vacuum Degeneracy and Conformal Mass in Lovelock AdS Gravity},
JHEP \textbf{11}, 128 (2017).
[arXiv:1710.08512]

\bibitem{Arenas2019}
G.~Arenas-Henriquez, R.~B.~Mann, O.~Miskovic and R.~Olea,
\textit{Mass in Lovelock Unique Vacuum gravity theories},
Phys. Rev. \textbf{D100}, 064038 (2019).
[arXiv:1905.10840]


\end{thebibliography}
\end{document}